\documentstyle[psfig,aps,twocolumn]{revtex}

\draft

\begin{document}

\wideabs{%

\title{Nonlinear sigma model study of a frustrated spin ladder}

\author{C.-M. Nedelcu$^{*}$, A. K. Kolezhuk$^{*,\dagger}$, and
	H.-J. Mikeska$^{*}$} 
\address{$^{*}$Institut f\"ur Theoretische Physik, Universit\"at
  Hannover, Appelstr. 2, D-30167 Hannover, Germany}
\address{$^{\dagger}$Institute of Magnetism, National Academy of Sciences and
Ministry of Education of Ukraine\\ Vernadskii avenue 36(b), 03142 Kiev,
Ukraine}

\date{\today}

\maketitle

\begin{abstract}
A model of two-leg spin-$S$ ladder with two additional frustrating
diagonal exchange couplings $J_{D}$, $J_{D}'$ is studied within the
framework of the nonlinear sigma model approach.  The phase diagram
has a rich structure and 
contains $2S$ gapless phase boundaries which split off the
boundary to the fully saturated ferromagnetic phase when $J_{D}$ and
$J_{D}'$ become different. For the $S={1\over2}$ case, the phase
boundary is identified as separating two topologically distinct
Haldane-type phases as discussed recently by Kim {\em et al.}
(cond-mat/9910023). 
\end{abstract}

}

\narrowtext

\section{Introduction}
\label{sec:intro}

Low-dimensional spin models still continue to attract a considerable
attention of researchers, both in theoretical and experimental
aspects. Since the famous prediction of Haldane in 1983
\cite{Haldane83} of different behaviour of Heisenberg spin chains with
integer and half-integer value of spin $S$, which was based on a
mapping to the nonlinear sigma model (NLSM), the NLSM approach was
recognized as an important tool in studying spin systems and have
found numerous applications (see, e.g., \cite{reviews} for a
review). Normally, the NLSM approach does not give good numerical
results, however it is usually able to capture the topology of the
phase diagram.

During the recent upsurge of interest to spin ladder models, several
researchers have successfully applied NLSM to describe a $N$-leg
spin-$S$ ladder \cite{Khveshchenko94,Senechal95,Sierra96}. In an
essential similarity to the case of a single chain, it was found that
for half-integer $S$, the ladders with even or odd number of legs $N$
are respectively gapped and gapless. A natural question arose, namely,
whether the properties of a gapped phase of, say, a two-leg
spin-$1\over2$ ladder are in some sense equivalent to those of the
Haldane phase of a spin-$1$ chain.

Several arguments were given in favour of the positive answer to the
above question.\cite{White96,Nishiyama+95,KolMik97} Particularly, it
was shown that by adding extra interactions one can introduce a
suitable generalization of the pure ladder model, increasing the
number of parameters in the phase space, and then one can find a path
in this generalized phase space which smoothly (i.e., without crossing
any phase boundaries) leads from the ladder model to a certain
composite representation of a spin-$1$ chain \cite{White96,KolMik97};
moreover, it was demonstrated \cite{White96,Nishiyama+95} that a
two-leg $S={1\over2}$ ladder has nonzero string order which is
believed \cite{string} to be a characteristic feature of the Haldane
phase.

On the other hand, it turned out that other generalizations may have
very different properties; for instance, for the model of a ``diagonal
ladder'' with additional equal-strength diagonal interactions, which
also yields a composite-spin representation of a spin-$1$
chain \cite{Xian95} one finds numerically that the ``usual'' ladder is
separated from the composite-spin-$1$ Haldane phase by a transition
line \cite{WKO98,Wang98}. An exactly solvable model exhibiting similar
features was also constructed \cite{KolMik98}.

Recently, Kim {\em et al.} \cite{Kim+99} have made an interesting
observation, noticing that there are actually at least {\em two}
different definitions of the string order for a two-leg spin-$1\over2$
ladder (depending on whether one combines the $S={1\over2}$ spins on
the rungs or on the diagonals). Exploiting the analogy with the
topological quantum numbers which can be introduced for short-range
valence bond states on a square lattice \cite{Bonesteel89}, they have
conjectured that those two definitions of the string order distinguish
between {\em two different Haldane-type phases}. This assumption was
supported by the results of the bosonization study of two
generalizations of the ladder model.

\begin{figure}
\mbox{\psfig{figure=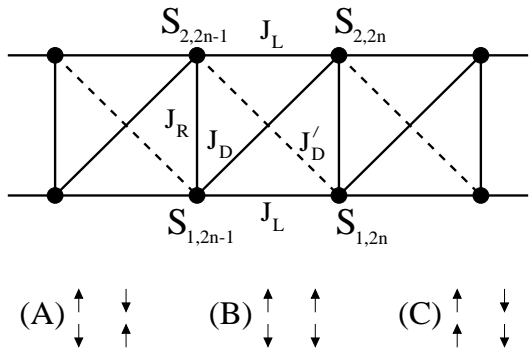,width=70mm}}
\vspace*{3mm}
  \caption{ A schematic representation of the generalized ladder model
  (\protect\ref{ham}). A, B, C denote different commensurate classical
  ground state configurations.}
  \label{fig:nlsm-lad}
\end{figure}

In hope to get a better understanding of the physics of the spin ladder,
and to search for possible new phase transitions, we find 
interesting to study the phase diagram of the generalized ladder model
with unequal diagonal couplings. We consider the model determined by
the Hamiltonian
\begin{eqnarray}
\label{ham}
\widehat H &=&J_{L}\sum_{\alpha=1,2}\sum_{i}
\mbox{\boldmath$S$\unboldmath}_{\alpha,i}
\mbox{\boldmath$S$\unboldmath}_{\alpha,i+1} 
+J_{R}\sum_{i}
\mbox{\boldmath$S$\unboldmath}_{1,i}
\mbox{\boldmath$S$\unboldmath}_{2,i} \nonumber\\
&+&\sum_{i}(J_{D}
\mbox{\boldmath$S$\unboldmath}_{1,i}
\mbox{\boldmath$S$\unboldmath}_{2,i+1}
+J_{D}'
\mbox{\boldmath$S$\unboldmath}_{2,i}
\mbox{\boldmath$S$\unboldmath}_{1,i+1})\,,
\end{eqnarray}
where $\mbox{\boldmath$S$\unboldmath}_{\alpha,i}$ are spin-$S$
operators at the $i$-th rung, $\alpha=1,2$ distinguishes the ladder
legs.  The model is schematically shown in Fig.\
\ref{fig:nlsm-lad}. At $J_{D}=J_{D}'=0$ one recovers a regular ladder,
while at $J_{D}'=0$ the model is equivalent to a zigzag chain with
alternation of the nearest-neighbour interaction. 
Interchanging $J_{D}$ and $J_{D}'$ is obviously equivalent to
interchanging the legs of the ladder, so that it is sufficient to
restrict ourselves to the $J_{D}\geq J_{D}'$ case. The point $J_{D}=J_{D}'$
is in a certain sense special, since it allows an additional symmetry
operation, namely, interchanging the spins on every other rung is then
equivalent to interchanging $J_{D}$ and $J_{L}$.

The phase space of the model is three-dimensional and is determined by
the three ratios of exchange constants, e.g., $J_{D}/J_{R}$,
$J_{D}'/J_{R}$, $J_{L}/J_{R}$.  We will show that for $J_{D}\not=
J_{D}'$ the phase diagram of the above model always possesses $2S$
gapless phase planes, which split off the boundary to the fully
saturated ferromagnetic phase at $J_{D}=J_{D}'$. We consider in detail
the most interesting case $S={1\over2}$ and show that the gapless
plane is an extension of one of the transition lines discussed in
Ref.\ \onlinecite{Kim+99}. In the next section we briefly describe the
mapping to NLSM, Sect.\ \ref{sec:discuss} contains the discussion of
the results, and, finally, Sect.\ \ref{sec:summary} gives a brief summary.

\section{Results of the mapping to the nonlinear sigma model}
\label{sec:nlsm}

To map the model (\ref{ham}) to a NLSM, we use the well-known
technique of spin coherent states path integral; this technique is
well described in reviews and textbooks\cite{reviews}, and here we
will not give a complete derivation but rather
indicate only the main steps.  We
choose a four-spin plaquette as an elementary magnetic cell; then
there are four classical ground states {\em commensurate} with this
choice of cell, namely a ferromagnetic state (F) and three modulated
states shown in Fig.\ \ref{fig:nlsm-lad} and denoted as (A), (B) and
(C). At $n$-th plaquette we introduce four variables ${\mathbf
m}_{n}$, ${\mathbf l}_{n}$, ${\mathbf u}_{n}$, ${\mathbf v}_{n}$,
defined as the following linear combinations of the ``classical'' spin
vectors (parameters of the coherent states):
\begin{eqnarray}
\label{ansatz}
{\mathbf l}_{n}&=&{1\over4S}({\mathbf S}_{1,2n-1}+{\mathbf S}_{2,2n-1}
-{\mathbf S}_{1,2n}-{\mathbf S}_{2,2n})\,,\nonumber\\
{\mathbf m}_{n}&=&{1\over4S}({\mathbf S}_{1,2n-1}+{\mathbf S}_{2,2n-1}
+{\mathbf S}_{1,2n}+{\mathbf S}_{2,2n})\,,\nonumber\\
{\mathbf u}_{n}&=&{1\over4S}({\mathbf S}_{1,2n}+{\mathbf S}_{2,2n-1}
-{\mathbf S}_{1,2n-1}-{\mathbf S}_{2,2n})\,,\\
{\mathbf v}_{n}&=&{1\over4S}({\mathbf S}_{2,2n-1}+{\mathbf S}_{2,2n}
-{\mathbf S}_{1,2n-1}-{\mathbf S}_{1,2n})\,,\nonumber
\end{eqnarray}
which satisfy the following four constraints:
\begin{eqnarray}
\label{constr}
&&{\mathbf m}^{2}+{\mathbf l}^{2}+{\mathbf u}^{2}+{\mathbf v}^{2}=1,
\nonumber\\
&&({\mathbf m}+{\mathbf l})\cdot({\mathbf u}+{\mathbf v})=0,\nonumber\\
&&({\mathbf m}-{\mathbf l})\cdot({\mathbf u}-{\mathbf v})=0,\\
&&({\mathbf m}+{\mathbf v})\cdot({\mathbf u}+{\mathbf l})=0,\nonumber 
\end{eqnarray}
Those variables we consider as smoothly varying functions of the space
coordinate $x_{n}=na$ when passing to the continuum limit; one should
mention that the above ansatz is essentially similar to that used by
S\'en\'echal \cite{Senechal95}. The advantage of the ansatz
(\ref{ansatz}) is that it conserves the total number of the degrees of
freedom, which is important to avoid ambiguities in the mapping, as
was recently realized on the example of inhomogeneous spin chains
\cite{Takano99}.

The order parameter for the four commensurate classical ground state
configurations F, A, B, C is respectively ${\mathbf m}$, ${\mathbf
u}$, ${\mathbf v}$, ${\mathbf l}$. 
Comparing the energies of those
configurations, one may obtain a ``draft'' of the classical phase
diagram which neglects presence of any incommensurate ground
states; for the moment
we are mainly interested in the commensurate
antiferromagnetic part, and the conditions for the existence of spiral
phases will be obtained later. One thus may treat ${\mathbf m}$ as a small
fluctuation, and obtain different field descriptions starting from one
of the configurations A, B, C. Massive degrees of freedom can be
integrated out in a usual way, and in each case one obtains the final
effective action in the form of a NLSM,
\begin{eqnarray}
\label{nlsm}
{{\cal A}_{\rm eff}/\hbar}&=&
{1\over2g}\int\int d\xi d\tau\, \{(\partial_{t} {\mathbf n})^{2}  
-(\partial_{x}{\mathbf n})^{2} \}\nonumber\\
&+&{\theta\over4\pi}\int\int d\xi d\tau\, 
{\mathbf n} \cdot (\partial_{x}{\mathbf n}\times \partial_{t}{\mathbf n})\,,
\end{eqnarray}
where ${\mathbf n}$ is the corresponding order parameter, and $\xi=x/a$,
$\tau=ct/a$ are dimensionless space-time variables, $a$ being the
lattice constant along the legs direction. For each of the
classical ``phases'' A, B, C the coupling
constant $g$ and the topological angle $\theta$ are given by the
following expressions:\\
{\em Phase A:} $J_{R}+2J_{L}>0$, $J_{D}^{+}<J_{R}$, $J_{D}^{+}<2J_{L}$.
\begin{eqnarray}
\label{resA}
&& g_{A}={J_{R}+2J_{L} \over 2S\sqrt{W_{A}}},\quad 
\theta_{A}=0\; \mbox{mod}\, 2\pi,\\
&& W_{A}={1\over4}(J_{R}+2J_{L})
\Big\{2J_{L}-J_{D}^{+}-{(J_{D}^{-})^{2}\over
J_{R}-J_{D}^{+}}\Big\}>0\,.
\nonumber
\end{eqnarray}
{\em Phase B:} $J_{R}+J_{D}^{+}>0$, $2J_{L}<J_{D}^{+}$, $2J_{L}<J_{R}$.
\begin{eqnarray}
\label{resB}
&& g_{B}={J_{R}+J_{D}^{+} \over 2S\sqrt{W_{B}}},\quad 
\theta_{B}={4\pi S J_{D}^{-}\over J_{R}+J_{D}^{+}},\\
&& W_{B}={1\over4}\Big\{(J_{R}+J_{D}^{+})
(J_{D}^{+}-2J_{L})-(J_{D}^{-})^{2}\Big\}>0\,.
\nonumber
\end{eqnarray}
{\em Phase C:} $J_{D}^{+}+2J_{L}>0$, $J_{R}<J_{D}^{+}$, $J_{R}<2J_{L}$.
\begin{eqnarray}
\label{resC}
&& g_{C}={J_{D}^{+}+2J_{L} \over 2S\sqrt{W_{C}}},\quad 
\theta_{C}=0\; \mbox{mod}\, 2\pi,\\
&& W_{C}={1\over4}(J_{D}^{+}+2J_{L})
\Big\{2J_{L}+J_{D}^{+}-{(J_{D}^{-})^{2}\over
J_{D}^{+}-J_{R}}\Big\}>0\,.
\nonumber
\end{eqnarray}
Here for the sake of convenience we have introduced the notations
\[
J_{D}^{\pm}\equiv (J_{D}\pm J_{D}')\,.
\]
The spin wave velocity for each case is given by
$c=2\sqrt{W}Sa/\hbar$. The inequalities define the boundaries of the
domains of validity of the correspondent mapping (not all of them
represent real phase boundaries, as will be discussed later).  The
boundaries defined by $W_{A,B,C}=0$  represent just the classical
conditions for the transition into a spiral phase. One may observe
that there is no spiral phase at $J_{D}^{-}=0$.

{\em Phase F} has to be considered separately, and it is
easy to obtain its
boundaries using the linear spin wave theory. There are two magnon
branches with the energies
\begin{eqnarray}
\label{ferro}
\varepsilon_{\pm}(q)&=&-S(J_{R}+J_{D}^{+}+2J_{L}) +2SJ_{L}\cos q
\nonumber\\
&\pm&S\Big\{ (J_{R}+J_{D}^{+}\cos q)^{2} +(J_{D}^{-}\sin
q)^{2}\Big\}^{1/2}\,, 
\end{eqnarray}
and from the condition of positiveness of $\varepsilon_{\pm}$ it is
easy to obtain the boundaries of the F phase. They are determined by
the inequalities
\begin{eqnarray}
\label{resF}
&&J_{R}+J_{D}^{+}>0,\quad J_{R}+2J_{L}>0,\nonumber\\
&& W_{F}\equiv -2J_{L}-J_{D}^{+}+(J_{D}^{-})^{2}/(J_{R}+J_{D}^{+}) >0\,.
\end{eqnarray}
At $W_{F}=0$, $\varepsilon_{-}(q)$ changes sign at once in a finite
interval of wave vectors near $q=0$, signaling the first-order transition,
$\varepsilon_{-}(q=\pi)$ vanishes at the line $J_{R}+2J_{L}=0$, and
$\varepsilon_{+}(q=0)$ becomes zero at the line $J_{R}+J_{D}^{+}=0$.

One can see that only in the (B) case there is a nontrivial
topological term, and the condition of gaplessness $\theta=(2n+1)\pi$
yields 
\begin{equation}
\label{gapless}
J_{D}^{-}={2n+1\over 4S}(J_{R}+J_{D}^{+}),\quad n=0,1,\ldots 2S-1\,.
\end{equation}
One can see that the $2S$ gapless planes (\ref{gapless}) exist only at
nonzero $J_{D}^{-}$, and at $J_{D}^{-}=0$ they split off the boundary
$J_{R}+J_{D}^{+}=0$ to the ferromagnetic phase.

\section{Discussion}
\label{sec:discuss}

\begin{figure}
\mbox{\psfig{figure=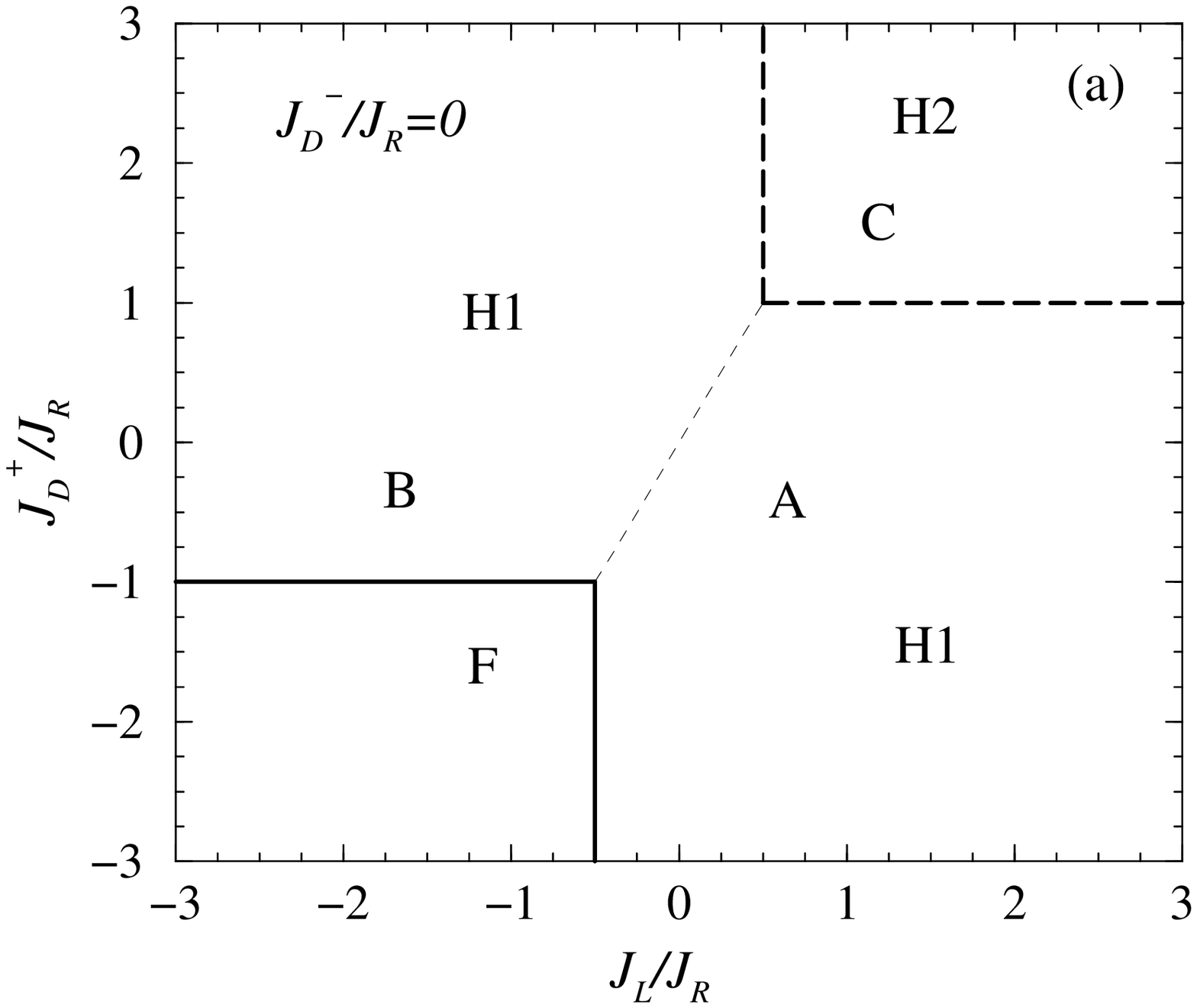,width=70mm}}
\vskip 1mm
\mbox{\psfig{figure=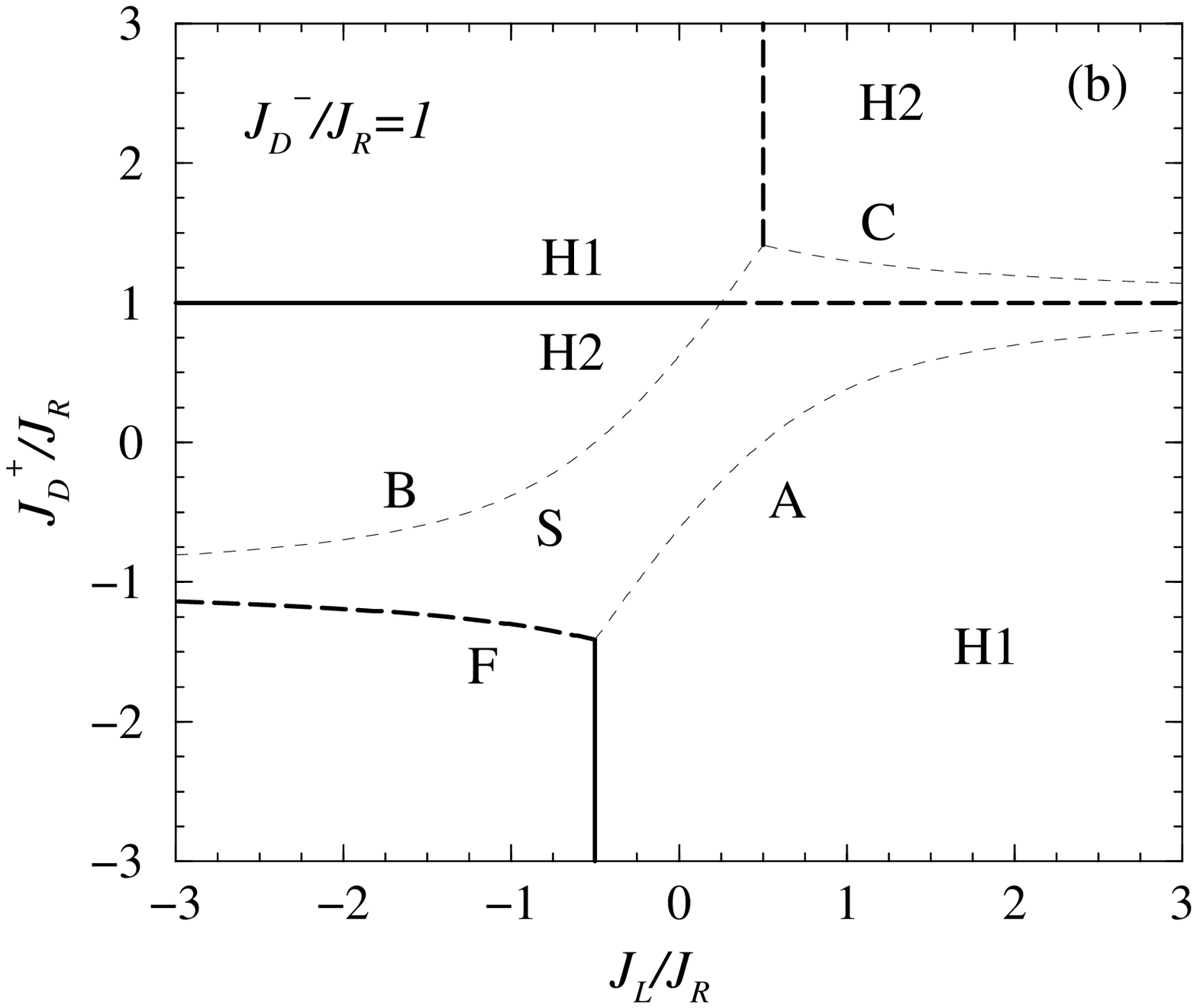,width=70mm}}
\vskip 1mm 
\mbox{\psfig{figure=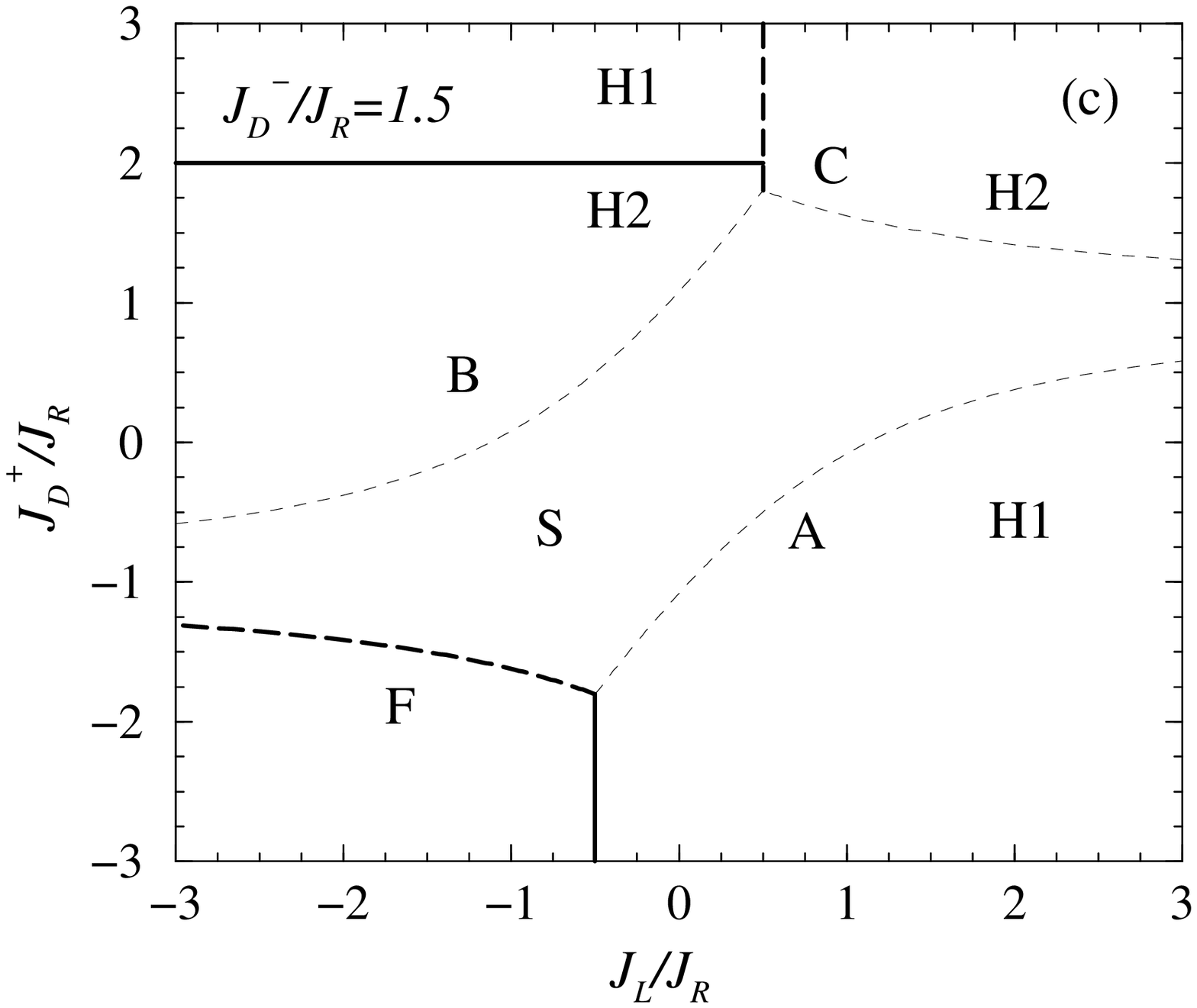,width=70mm}} 
\caption{ A sketch of the phase diagram of the model
(\protect\ref{ham}), shown are the slices in the three-dimensional
phase space at fixed $J_{D}^{-}\equiv J_{D}-J_{D}'$: (a)
$J_{D}^{-}=0$; (b) $J_{D}^{-}=J_{R}$; (c) $J_{D}^{-}=3J_{R}/2$. Thick
solid and dashed lines denote the second- and first-order transition
boundaries, respectively. Thin dashed lines indicate crossover between
different classical configurations; the coupling constant diverges at
those lines. A, B, C are classical configurations shown in Fig.\
\protect\ref{fig:nlsm-lad}, F denotes the fully saturated
ferromagnetic phase, and S stands for the spiral ``phase'' inside
which our approach is not valid.  H1 and H2 denote topologically
different Haldane-type phases with ${\cal O}_{\rm even}\not=0$ and
${\cal O}_{\rm odd}\not=0$ according to the classification of Ref.\
\protect\onlinecite{Kim+99}.}
\label{fig:pd}
\end{figure}

Let us concentrate on the case $S={1\over2}$ as being the most
important one. For $S={1\over2}$, a sketch of the resulting phase
diagram is presented in Fig.\ \ref{fig:pd} in a form of
two-dimensional slices through the phase space at three fixed values of
$J_{D}^{-}$ ($J_{R}$ is considered to be positive). 

At $J_{D}^{-}=0$ there are no other gapless lines except the
boundaries of the ferromagnetic phase, and there is no spiral phase.
The coupling constants $g_{A}$, $g_{B}$ diverge at the (AB) boundary
$J_{D}^{+}=2J_{L}$, which indicates that this classical phase boundary
gets destroyed by quantum fluctuations. On the other hand, all the
coupling constants  remain finite at the (BC) and (AC) boundaries, but
they undergo a jump when crossing the boundaries, which suggests a
first-order transition. This is in agreement with the numerical
\cite{WKO98,Wang98} and bosonization \cite{WKO98,Kim+99} studies,
showing the presence of a first-order transition with $J_{D}=J_{R}/2$
being the asymptote for the transition line at $J_{L}\to\infty$. 
There is also
a ``mirror'' transition line $J_{L}=J_{R}/2$ due to the
$J_{D}\leftrightarrow J_{L}$ symmetry. According to the
classification of Ref.\ \onlinecite{Kim+99}, those two first-order
transition lines separate two topologically different Haldane-type
phases with ${\cal O}_{\rm even}\not=0$ (phases A,B) and ${\cal
O}_{\rm odd}\not=0$ (phase C); below we refer to those two phases as
H1 and H2 (see Fig.\ \ref{fig:pd}(a)).

At finite $J_{D}^{-}$, the spiral phase (S) appears classically in a
finite region of the phase diagram. The S region is for us just a ``white
spot'' which cannot be treated within the present approach; to
construct an effective description for the (S) phase, one has to
employ different techniques.   At finite
$J_{D}^{-}$ the gapless line $J_{D}^{+}=2J_{D}^{-}-J_{R}$ starts to
split off the (BF) boundary $J_{D}^{+}=J_{R}$, and the (FS) boundary
becomes first order, as one sees from the behaviour of
$\varepsilon_{-}(q)$ (cf. (\ref{ferro})). The coupling constants $g_{A,B,C}$
diverge at the boundaries to the S phase, which suggests
destruction of any (quasi)-long-range order, and thus one may expect
that the gapless line terminates at the (BS) boundary, though it may
in principle continue as a first-order transition line.

It is worthwhile to look at the particular case
$J_{D}^{-}=J_{R}$. According to Ref.\ \onlinecite{Kim+99}, the gapless
line $J_{D}^{+}=J_{R}$ in this case also separates two phases with
different string order, which implies that the lower portion of the B
phase belongs to the H2 class. It is also known that in this case the
gapless line continues at larger $J_{L}$ as the first-order line
(recall that $J_{D}^{-}=J_{D}^{+}=J_{R}$ corresponds to the uniform
spin chain with next-nearest neighbour interaction). Thus, it becomes
clear that additional phase boundaries should exist somewhere inside
the spiral ``phase,'' to achieve a proper separation of H1- and
H2-type phases (see Fig.\ \ref{fig:pd}(b,c)). This could be an
interesting topic for the future work.

\section{Summary}
\label{sec:summary}

We have studied the phase diagram of the generalized ladder model with
unequal diagonal couplings $J_{D}$, $J_{D}'$ within the framework of
the nonlinear sigma model. We show that the phase diagram has a rich
structure including several first- and second-order transition
boundaries. There exist $2S$ gapless phase boundaries which split off
the boundary to the ferromagnetic phase at $J_{D}\not=J_{D}'$. We
consider the case $S={1\over2}$ in more detail and show that the
gapless plane is an extension of one of the transition lines discussed
in Ref.\ \onlinecite{Kim+99} which separate Haldane-type phases with
different topological order parameter. Still, several features of the
phase diagram remain unclear and require further study.

\acknowledgements

This work was supported by the German Federal Ministry for Research
and Technology (BMBFT) under the contract 03MI5HAN5.  A.K.  gratefully
acknowledges the hospitality of the Hannover Institute for Theoretical
Physics. C.N. was supported by the DFG-Graduiertenkolleg ``Quantum Field
Theory Methods in Particle Physics, Gravity, and Statistical Physics''.

\end{document}